\documentclass{cplslarge}%
\usepackage[authoryear]{natbib}
\usepackage{makeidx}

\usepackage[switch]{lineno}

\usepackage[bookmarks = true, bookmarksnumbered = true, pdfpagemode =None, pdfstartview = FitH, pdfpagelayout = SinglePage, colorlinks = true, urlcolor = red, citecolor = blue]{hyperref}

\usepackage{graphics}

\makeindex

\global\chapterreferencefalse
\usepackage[sectionbib]{chapterbib}

\begin{document}
\frontmatter

\mainmatter

\author[Cao, Dougherty, Hunt, Bunce, Christensen, Khurana, \&\ Kivelson]{h. cao$^{1,2,3}$, m.k. dougherty$^3$, g.j. hunt$^3$, e.j. bunce$^4$, u.r. christensen$^5$, k.k. khurana$^1$, and m.g. kivelson$^1$}
\chapter{Saturn's Magnetic Field at Unprecedented Detail Achieved by Cassini's Close Encounters}

\footnotesize
$^1$ Department of Earth, Planetary, and Space Sciences, University of California, Los Angeles, Charles Young Drive East, Los Angeles, CA 90095, USA. \\
$^2$ Department of Earth and Planetary Sciences, Harvard University, 20 Oxford Street, Cambridge, MA 02138, USA.\\
$^3$ Physics Department, The Blackett Laboratory, Imperial College London, London, SW7 2AZ, UK.\\
$^4$ Department of Physics and Astronomy, University of Leicester, Leicester, LE1 7RH, UK.\\
$^5$ Max Planck Institute for Solar System Research, Justus-von-Liebig-Weg 3, 37077 Göttingen, Germany.

\section*{Copyright Notice}
The Chapter, ``Saturn's Magnetic Field at Unprecedented Detail Achieved by Cassini's Close Encounters'', is to be published by Cambridge University Press as part of a multi-volume work edited by Kevin Baines, Michael Flasar, Norbert Krupp, and Thomas Stallard, entitled ``Cassini at Saturn: The Grand Finale" (`the Volume')
 
\copyright  ~in the Chapter, H. Cao, M.K. Dougherty, G.J. Hunt, E.J. Bunce, U.R. Christensen,  K.K. Khurana and M.G. Kivelson,

\copyright ~in the Volume, Cambridge University Press
 
NB: The copy of the Chapter, as displayed on this website, is a draft, pre-publication copy only. The final, published version of the Chapter will be available to purchase through Cambridge University Press and other standard distribution channels as part of the wider, edited Volume, once published. This draft copy is made available for personal use only and must not be sold or re-distributed.

\normalsize

\section*{Abstract}
The last 22.5 orbits of the Cassini mission brought the spacecraft to less than 3000 $km$ from Saturn's 1-bar surface. These close encounters offered an unprecedented view of Saturn's magnetic field, including contributions from the internal dynamo, the ionosphere, and the magnetosphere. In this chapter, we highlight the new picture of Saturn's magnetic field from the Cassini mission including {the persistent yet time-varying low-latitude field-aligned currents, Alfvén waves planetward of the D-ring, extreme axisymmetry, and high-degree magnetic moments}. We then discuss the implications and new questions raised for Saturn's {innermost magnetosphere, equatorial ionosphere, and interior}. We conclude this chapter with an outlook for the future exploration of Saturn and other giant planets. 

\section{Introduction}\label{sec:intro}

{Magnetic field investigations have played central roles in our multi-disciplinary exploration of the Saturn system \citep[e.g.,][]{Dougherty2006}}. The intrinsic magnetic field, originating in the deep interior, determines key properties of the electromagnetic environment around planet Saturn and its rings and moons. The Grand Finale phase of the Cassini mission (Apr. to Sep. 2017) brought the spacecraft to extreme proximity to Saturn, its trajectory passing through the gap between its upper atmosphere and its innermost ring, a region never visited before. The Cassini Grand Finale (CGF) orbits provided an unprecedented opportunity to decipher Saturn's interior, ionosphere, and the innermost magnetosphere. \index{magnetic field investigation} \index{interior}

\begin{figure*}
\begin{center}
\figurebox{5.5in}{}{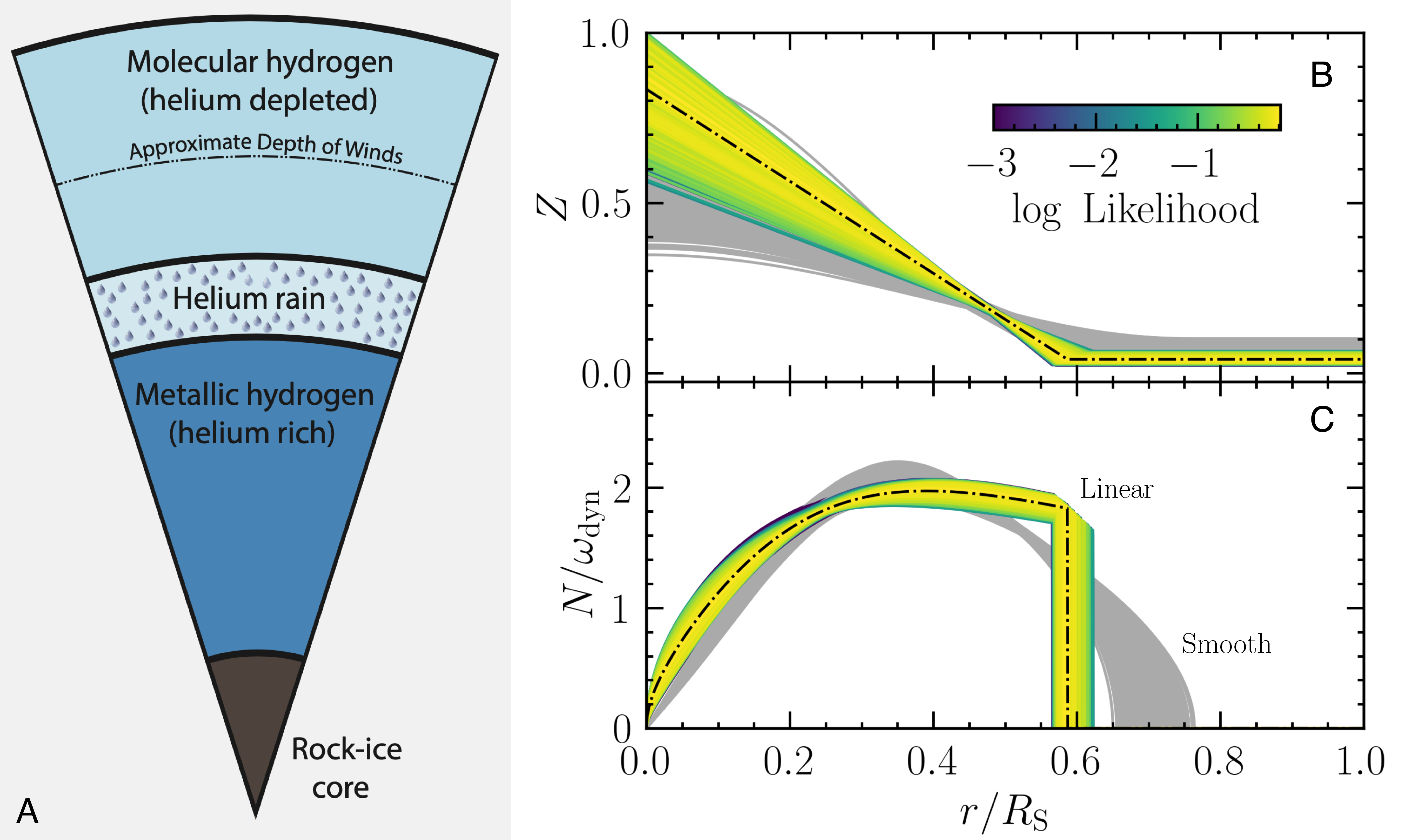}
\caption{Pictures of Saturn's bulk interior structure. Panel A shows a classical four-layer model of Saturn's interior structure constructed to match Saturn's gravity field: a molecular hydrogen layer, a helium rain layer, a metallic hydrogen layer, and a rock-ice core. Panels B \& C depict a new picture of Saturn's interior with a large diffuse core extending to about 0.6 Saturn radii. In this new picture, the heavy element ($Z$) abundance features a gradual transition from the center of Saturn to about 0.6 Saturn radii (panel B). The diffuse core of Saturn is stably stratified, featuring Brunt-Väisälä frequencies $N$ as high as twice that of the natural frequency of Saturn $\omega_{dyn}=(GM_S/R_S^3)^{1/2}$ where $G$ is the gravitational constant, $M_S$ is the mass of Saturn, and $R_S$ is the radius of Saturn. Panel A is from \citet[]{militzer2019}, while Panels B \& C are from \citet[]{MankovichFuller2021}.}
\label{fig:SaturnInterior}
\end{center}
\end{figure*}

{In the conventional picture, the bulk interior of Saturn is} qualitatively divided into four layers based on composition or material properties \citep[]{militzer2019}: a central rock-icy core, a metallic hydrogen layer \citep[]{weir1996}, a helium rain layer \citep[]{Morales2009, Brygoo2021}, and the outer molecular hydrogen layer {(e.g., see Fig. \ref{fig:SaturnInterior}A). Given our current understanding of material properties in the relevant pressure-temperature ranges, the transition between each adjacent pair of layers likely is smooth \citep[]{weir1996, WM2012}. Employing ring seismology in addition to gravity, \citet[]{MankovichFuller2021} showed that a diffusive, stably-stratified core could extend to $\sim$ 0.6 Saturn radii (Fig. \ref{fig:SaturnInterior}BC). The origin (e.g., giant impact versus core erosion) and dynamical consequences (e.g., flow characteristics and magnetic field generation) of such an extended diffusive core inside Saturn are active areas of research.} \index{interior} \index{metallic hydrogen} \index{helium rain} \index{ring seismology} \index{gravity} \index{stably stratified} \index{diffusive core}

Convective motion, {either in the form of overturning convection or double diffusive convection \citep[]{Garaud2018}}, is primarily driven by the cooling of the planet and the subsequent phase-separation (e.g., helium rain) \citep[]{Stevenson1980}. Magnetic field generation is a natural outcome of rapidly rotating {overturning} convection in electrically conducting fluids, a process commonly referred to as magnetohydrodynamic (MHD) dynamo action \citep[e.g.,][]{moffatt_dormy_2019}. {Whether double diffusive convection can support planetary-scale magnetic field generation still needs to be clarified.} Measuring the characteristics of the intrinsic magnetic field, including its time variations, would provide {key observational constraints}. \index{dynamo} \index{double diffusive convection}

Another common feature of rapidly rotating convection is the emergence of differential rotation, also referred to as zonal flows or jets, in which different parts of the fluid feature different mean angular velocities with respect to the spin-axis. The zonal flows on Saturn's surface feature equatorial super-rotation that are about 4\% faster than the bulk rotation, and alternating bands of super-rotation and sub-rotation in the off-equatorial region \citep[]{Read2009}. Deep zonal flows will inevitably interact with the planetary magnetic field at depths with even modest electrical conductivity \citep[]{liu2008, cao2017b}. For example, the Lorentz force associated with the interaction could play a role in truncating the zonal flow \citep[]{Christensen2020, GastineWicht2021} and the zonal flow could modify the magnetic field \citep[]{cao2017b}. \index{rapidly rotating convection} \index{differential rotation} \index{zonal flow} \index{super-rotation} \index{angular velocity}

The bulk characteristics of Saturn's ionosphere have been constrained with radio occultation and ground-based $H_3^+$ emission measurements \citep[]{Kliore2014, Miller2020}. These measurements revealed a highly dynamic ionosphere. In addition to the expected strong $H_3^+$ emission at auroral latitudes, the mid-to-low latitude $H_3^+$ emission at Saturn shows distinct patterns correlated with the spatial structure of the rings when mapped along magnetic field lines \citep[]{OD2013}. This led to the suggestion of material in-fall from the rings along magnetic field lines to the ionosphere of Saturn, a phenomenon sometimes referred to as ``ring rain" \citep[]{CW1984}. The material exchange and electromagnetic interaction between Saturn and its rings likely are defining factors of the innermost magnetosphere at Saturn. Moreover, these interactions could shape the long term evolution of Saturn's rings. \index{ionosphere} \index{ring rain}

In section \ref{sec:pre-CGF} we present a brief overview of pre-Cassini Grand Finale knowledge of Saturn's magnetic field. In section \ref{sec:CGF}, we summarize the new picture of Saturn's magnetic field revealed by the Cassini Grand Finale. In section \ref{sec:implications}, we discuss the implications for Saturn's innermost magnetosphere, ionosphere, and interior as well as open questions. In section \ref{sec:outlook}, we offer our outlook for future exploration of Saturn and other giant planets. 

\adjustfigure{135pt}

\section{A brief overview of pre-Cassini Grand Finale knowledge of Saturn's magnetic field}\label{sec:pre-CGF}

Saturn's magnetic field was discovered during the Pioneer 11 Saturn flyby. Together with data from the subsequent two flybys by Voyager 1 \& 2 , these measurements revealed that Saturn's large-scale internal magnetic field is dominated by the axial dipole, with $\sim$10\% contributions from {axial quadrupole and octupole} magnetic moments when evaluated on the 1-bar surface \citep[]{DavisSmith1990}.  Saturn's surface magnetic field strength is slightly weaker than that of the Earth and almost one order of magnitude weaker than that of Jupiter. This relatively weak surface field at Saturn was a surprise given the relatively strong surface heat flux measured at Saturn \citep[]{INGERSOLL79}. A bigger surprise of Saturn's internal magnetic field was the seeming lack of departure from symmetry around the spin-axis, an unexpected feature based on a naive interpretation of Cowling's theorem \citep{Cowling1933} which precludes a purely axisymmetric magnetic field from being maintained by dynamo action. The sparse spatial coverage of these three flybys did not yield a stringent upper-limit on the departure from axisymmetry in Saturn's internal magnetic field: a dipole tilt on the order of 1$^\circ$ was still permitted by these early measurements. \index{Pioneer} \index{Voyager} \index{Cowling's theorem} \index{dipole tilt} \index{axisymmetric magnetic field} \index{dynamo} \index{magnetic moment}

{The Cassini magnetometer measurements in the first few years offered reasonably good latitude-longitude coverage for investigating non-axisymmetry in Saturn's internal magnetic field. \citet{SterenborgBloxham2010} analyzed the Cassini magnetometer data between Jun 2004 and May 2008 within 3.9 $R_S$ and placed an upper bound on the order-1 non-axisymmetric magnetic moments to be less than $\sim$ 5\% of the total axisymmetric moments. \citet[]{Cao2011} restricted their analysis to measurements inside the magnetic shell connecting Saturn and the orbit of Enceladus between 2005 and 2010 and} placed stringent constraints on the low-degree non-axisymmetry in Saturn's internal magnetic field: the dipole tilt must be smaller than 0.06$^\circ$ while the non-axisymmetric quadrupolar moments are less than 6\% of the axisymmetric quadrupolar moment. The non-detection of internal magnetic non-axisymmetry also implies that the rotation rate of the deep interior of Saturn needs to be determined from other means {such as gravity, shape, wind shear, and ring seismology} \citep[]{militzer2019, Mankovich2019}. \index{non-axisymmetry} \index{dipole tilt} \index{rotation rate} \index{gravity} \index{ring seismology} \index{magnetic moment} \index{Enceladus}

One important measure of a planetary internal magnetic field is its change with time \citep[e.g.,][]{HolmeOlsen2006}, commonly referred to as the magnetic secular variation (SV), which offers a window into the internal dynamo flows and waves. \citet[]{Cao2011} compared their low-degree model of Saturn's internal magnetic field built from the pre-Grand Finale Cassini measurements to the SPV {(Saturn Pioneer Voyager)} model \citep[]{DavisSmith1990} built from the Pioneer 11, Voyager 1 \& Voyager 2 measurements, and found that the differences in Saturn's low-degree internal magnetic moments between the Pioneer-Voyager era and the early Cassini era are very small. If interpreting the difference (with their uncertainty ranges) as the linear secular variation rate, then Saturn's magnetic SV rate is on the order of 1 $nT/year$ or less which is much smaller than the order 10 $nT/year$ secular variation rate of the recent geomagnetic field \citep[]{Finlay2020}. \index{magnetic secular variation} \index{dynamo} \index{Pioneer} \index{Voyager} \index{magnetic moment}

\section{A new picture of Saturn's magnetic field from the Cassini Grand Finale measurements}\label{sec:CGF}

Following a gravity assist from Titan, the Cassini spacecraft embarked upon the Grand Finale phase of its journey, which {consisted} of 22 high inclination orbits with periapses in the gap between Saturn and its innermost ring, before descending into the atmosphere of Saturn on 15 Sep. 2017. The periapsis altitudes of the Cassini Grand Finale (CGF) orbits ranged between 3911 $km$ and 1444 $km$ from Saturn's 1-bar surface \citep[e.g., see Table 1 in][]{Cao2019}, significantly lower than the altitudes of all previous Cassini orbits and Pioneer-Voyager flybys. For comparison, the periapsis altitudes of Cassini SOI and Pioneer 11 were $\sim$ 20,000 $km$. \index{Pioneer}

\begin{figure}%
\begin{center}
\figurebox{3.25in}{}{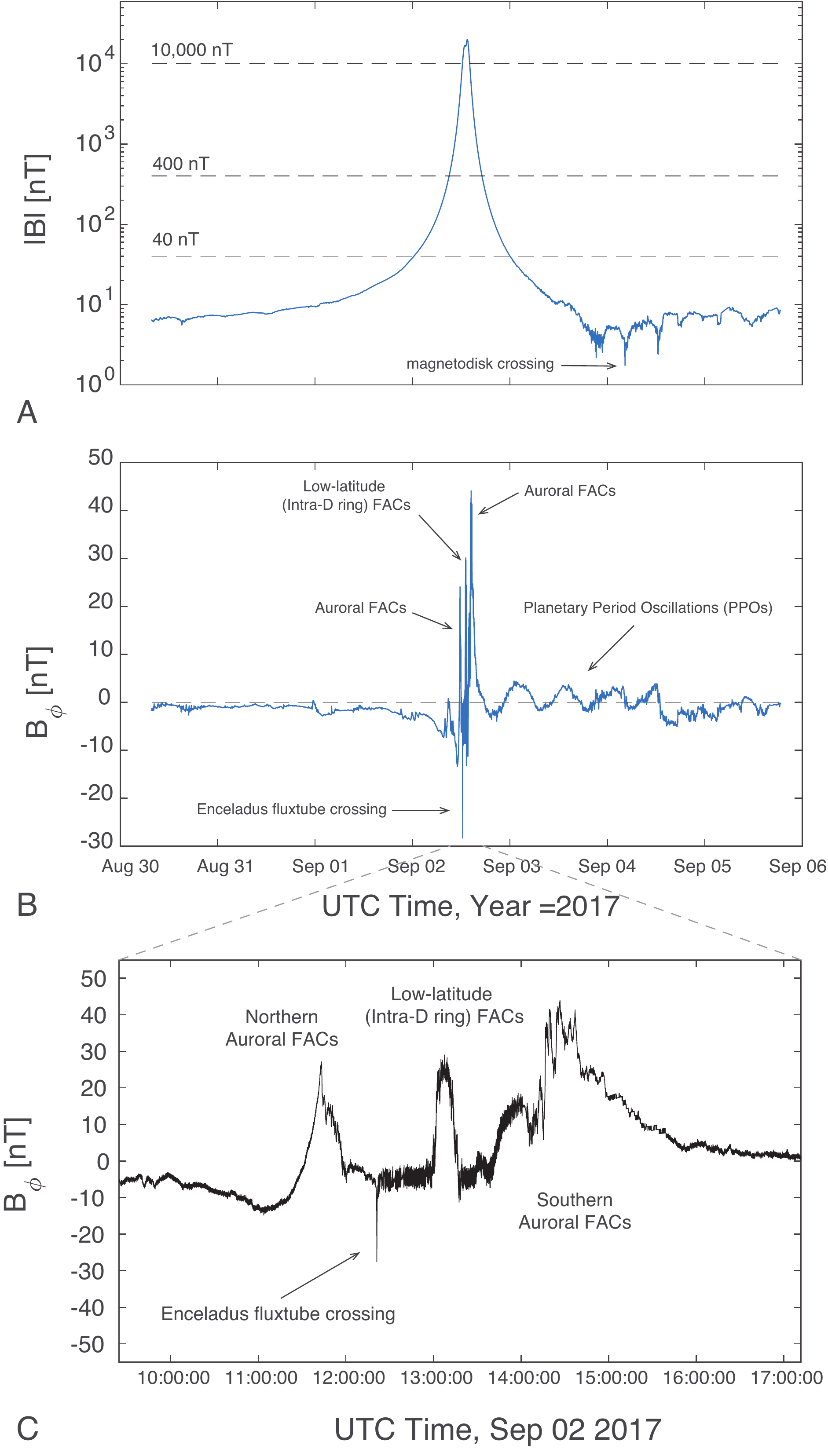}
\caption{Magnetic field strength and the azimuthal component measured during a 6.5-day Cassini Grand Finale orbit. Panel A shows the total field strength on a log-scale, while Panel B shows the azimuthal field component $B_\phi$ on a linear scale. Panel C shows a zoom-in of $B_\phi$ for a few hours around the periapsis. Figure from \citet[]{Cao2019}.}
\label{fig:B_Bphi_Rev291}
\end{center}
\end{figure}

In an inertial frame, the periapses of the CGF orbits were near local noon while the apoapses were near {midnight}. The inclination of the CGF orbits, with respect to the rotational equator of Saturn, was $\sim$ 62$^\circ$. Figure \ref{fig:B_Bphi_Rev291} displays the measured magnetic field strength $|B|$ and the azimuthal component, $B_\phi$, along a 6.5-day CGF orbit (Rev 291), which serves as an illustration of the bulk features of Saturn's magnetic field. It can be seen that the magnetic field strength varied by 4 orders of magnitude, from $<$ 2 $nT$ to $>$ 20,000 $nT$, during a CGF orbit. The minimum field strength was recorded during one crossing of the magnetodisk \citep[]{Bunce2007}. Multiple crossings or close approaches to the magnetodisk are evident from the periodic decreases of the magnetic field strength during the outbound portion of the pass. The azimuthal component, which stayed within $[-30, 50] \; nT$ while the background field varied by 4 orders of magnitude, revealed many dynamical features of Saturn's magnetosphere. A northern-hemisphere high-latitude crossing of the magnetic field lines connecting Saturn and Enceladus occurred during this orbit, see feature labelled as Enceladus flux tube crossing on 02 Sep. 2017, where a strong negative $B_\phi$ ($\sim$ $-30$  $nT$) was observed. The sign of $B_\phi$ there is consistent with a bend-back of Saturn's field lines resulting from the interaction of flowing magnetospheric plasma with Enceladus \citep[]{Dougherty2006}, and the amplitude of the associated field-aligned currents (FACs) has been estimated to be $\sim$ 200 $nA \, m^{-2}$ \citep[]{Sulaiman2018a}. The expected auroral FACs \citep[]{Hunt2020} and the ubiquitous planetary period oscillations (PPOs) \citep[]{Provan2018,Provan2019c} are also evident in the measured $B_\phi$ component. {\cite{Provan2019c} showed that the dual modulations by the northern and southern PPOs are present throughout Saturn's innermost magnetosphere in all three magnetic field components.} Moreover, the measured $B_\phi$ revealed a new feature in Saturn's innermost magnetosphere: a low-latitude FAC system located near the magnetic shells connecting Saturn and its tenuous D-ring \citep[]{DC2018,Khurana2018,Provan2019b,Hunt2019}. \index{azimuthal magnetic field} \index{magnetodisk} \index{Enceladus} \index{field-aligned current} \index{FAC} \index{planetary period oscillations} \index{PPO} \index{flux tube} \index{D-ring}

\subsection{Discovery of a low-latitude field-aligned current system}
\label{sec:Bphi_obs}

\begin{figure}%
\begin{center}
\figurebox{3.25in}{}{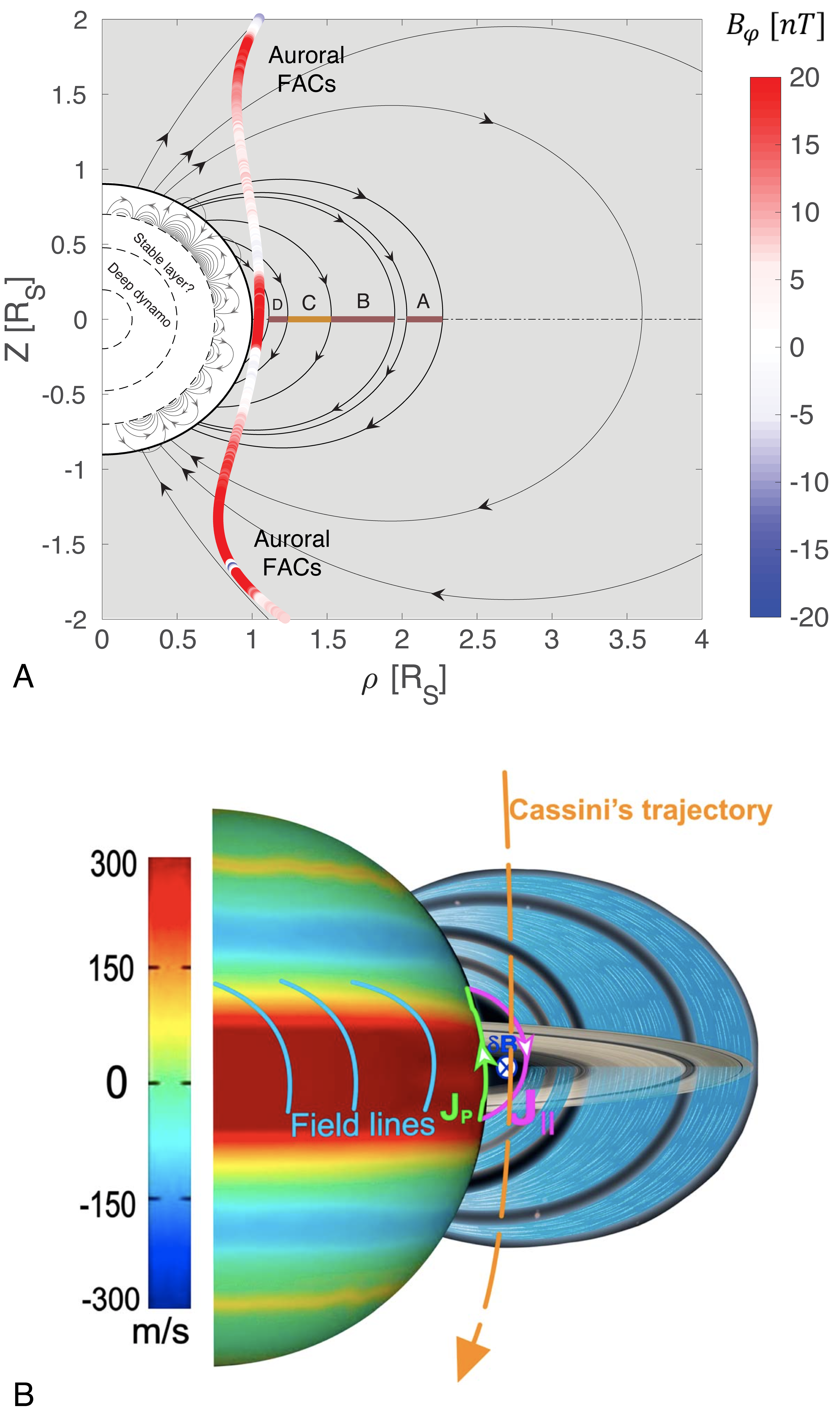}
\caption{Panel A shows the measured azimuthal magnetic field, $B_\phi$, around the periapsis of the first Cassini Grand Finale orbit (Rev 271) color-coded along the trajectory of the orbit. The high-latitude auroral FACs and the low-latitude intra-D ring FACs are both evident from the measured $B_\phi$. Panel B shows a sketch of the magnetospheric-ionospheric current system consistent with the observed positive $B_\phi$ in Saturn's innermost magnetosphere. {$J_P$ is the Pedersen current while $J_\parallel$ is the field-aligned current.} Panel A is from \citet[]{DC2018} while Panel B is from \citet[]{Khurana2018}.}
\label{fig:Dougherty2018_Fig0}
\end{center}
\end{figure}

\begin{figure*}%
\begin{center}
\figurebox{5.5in}{}{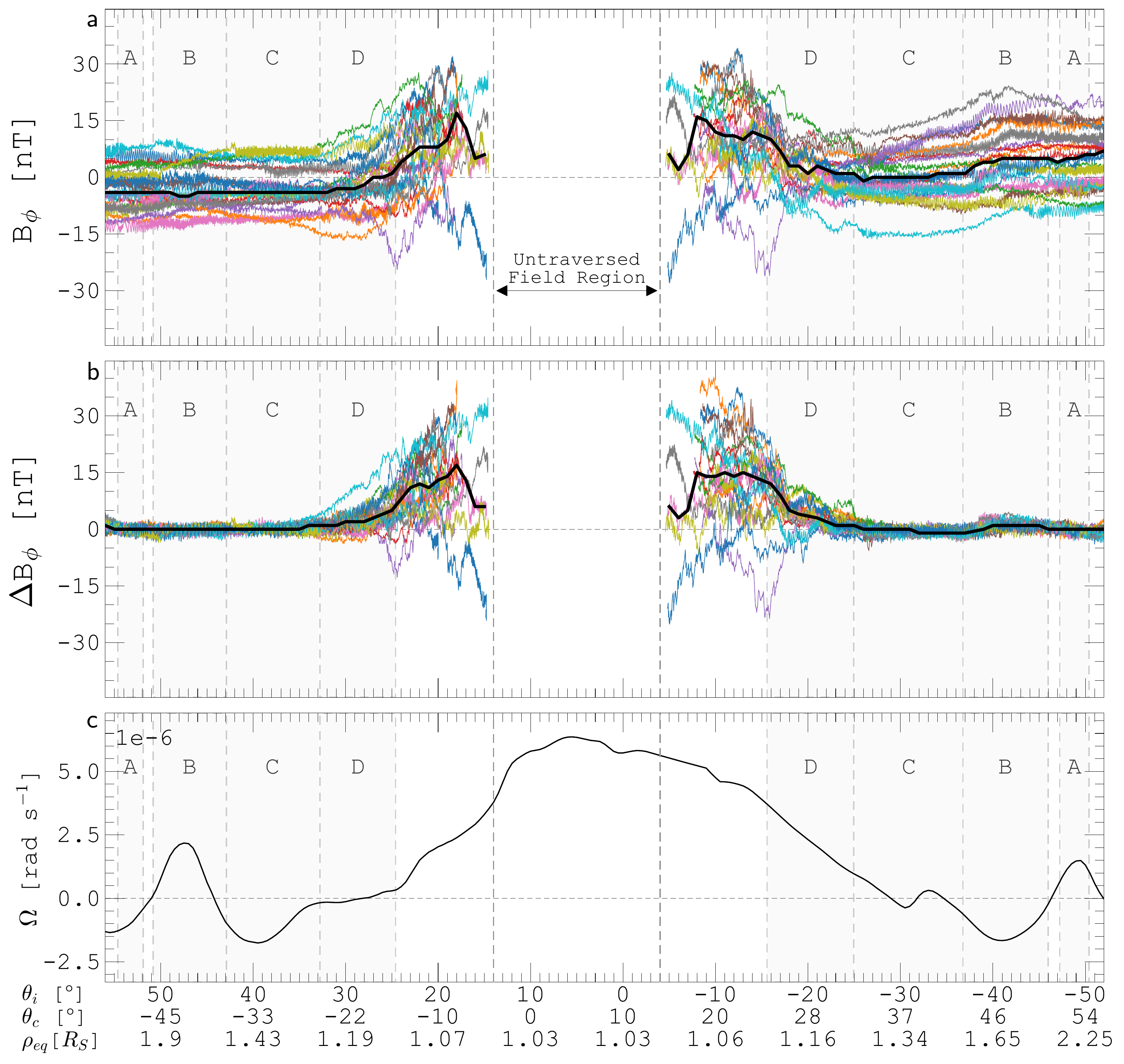}
\caption{The measured azimuthal magnetic field, $B_\phi$, along the Cassini Grand Finale orbits. {The horizontal axis is labelled with the mapping of the magnetic field line traversed by the spacecraft to the ionosphere and the ring plane: $\theta_i$ is the ionospheric latitude of the field line footprint in the same hemisphere as the measurement, $\theta_c$ is the ionospheric latitude of the conjugate footprint in the opposite hemisphere along the field line, $\rho_{eq}$ is the (cylindrical) radial distance in the equatorial/ring plane of the field line from the center of Saturn.} The top panel shows the total observed $B_\phi$ while the middle panel shows de-trended {field $\Delta B_\phi$} where a fourth order polynomial fit has been subtracted from the data to isolate the intra-D ring features. {In both panels, colored traces represent different CGF orbits while the black traces represent the running median from all orbits. The bottom panel shows the angular velocity profile of the 1-bar zonal winds at Saturn as a function of $\theta_i$,} referenced to an assumed planetary rotation period of 10 hr 34 min 13 sec. Figure from \citet[]{Agiwal2021}.}
\label{fig:Bphi_from_CGF}
\end{center}
\end{figure*}

Unexpected $B_\phi$ perturbations on the order of 20 $nT$ were consistently observed around periapses along CGF orbits \citep[]{DC2018}. When mapped along Saturn's background magnetic field lines, these $B_\phi$ perturbations map to the tenuous D-ring of Saturn and its planetward cavity (see Figs. \ref{fig:Dougherty2018_Fig0} \& \ref{fig:Bphi_from_CGF}). The sharpness of the spatial variation leading to the center peak is a strong indicator that these $B_\phi$ signatures are of magnetospheric-ionospheric origin, instead of a deep interior origin. The predominantly positive nature of the low-latitude $B_\phi$ perturbations (see Fig. \ref{fig:Bphi_from_CGF}) further points to their toroidal nature, which must be associated with local meridional electric currents. \index{D-ring}

The electric current system associated with positive low-latitude $B_\phi$ is inter-hemispheric, with field-aligned currents {$J_\parallel$} flowing from the north to the south in the magnetosphere and return Pedersen currents {$J_P$} in the ionosphere flowing south to north as illustrated in Fig. \ref{fig:Dougherty2018_Fig0}B. {A negative $B_\phi$ would correspond to a reversed current loop.} \citet[]{Khurana2018} further proposed that such a low-latitude electric current system is driven by zonal wind shear between the two ``ends" of Saturn's magnetic field embedded in the ionosphere. \citet[]{Khurana2018} estimated the Joule dissipation associated with this low-latitude current system and found it to be $\sim$ $2 \times 10^{11}$ $W$, assuming a height-integrated conductance $\sim$ 9 $S$ (siemens). This dissipation rate is similar to the estimated heating rate from solar extreme ultraviolet radiation at Saturn, but is not sufficient to explain the higher-than-expected temperature in Saturn's equatorial thermosphere \citep[]{MW2006}. \index{Pedersen current} \index{zonal wind shear} \index{Joule dissipation} \index{thermosphere} \index{inter-hemispheric} \index{field-aligned current} \index{ionosphere}

\begin{figure}
\begin{center}
\figurebox{3.25in}{}{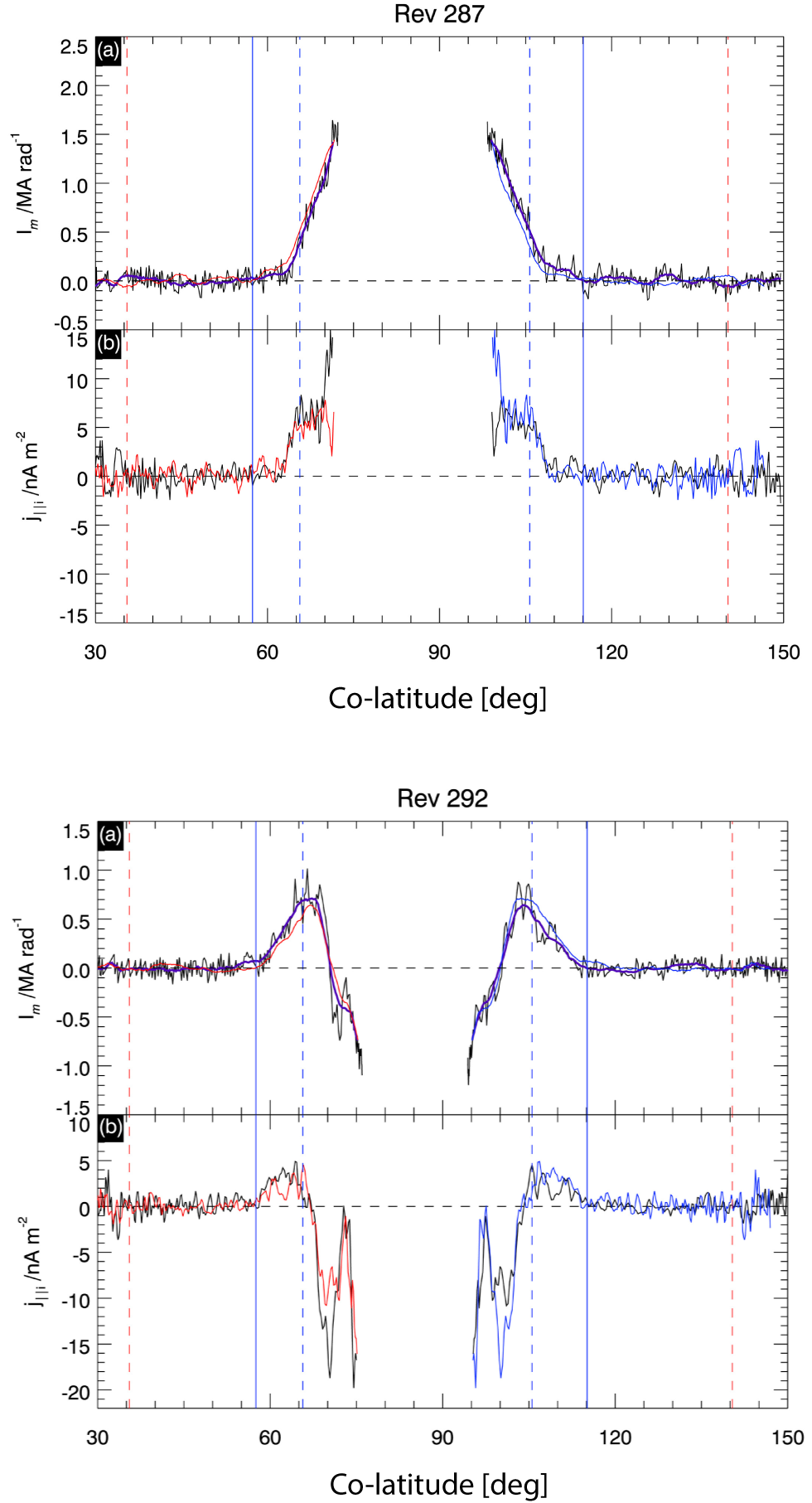}
\caption{Ionospheric meridional current $I_m$ and field-aligned current density $j_{\parallel \, i}$ associated with the low-latitude $B_\phi$ observed along Cassini Rev 287 (top) and Rev 292 (bottom). The vertical lines indicate magnetic mapping to ring boundaries: the pair of red dashed lines correspond to the outer boundary of the A-ring, while the solid (dashed) pair of blue lines correspond to the outer (inner) boundaries of the D-ring. Figure from \citet[]{Hunt2019}.}
\label{fig:Bphi_FAC}
\end{center}
\end{figure}

\citet[]{Hunt2019} estimated the electric current density associated with these low-latitude $B_\phi$ perturbations. Assuming that the azimuthal extent of the current system is much wider than that traversed by the spacecraft, they inferred that the low-latitude ionospheric meridional currents are $\sim$ 0.5 - 1.5 $MA$ (Mega Ampere) per radian of azimuth (see Fig. \ref{fig:Bphi_FAC} {for two examples}), similar in intensity to the auroral region current \citep[e.g.,][]{Hunt2020}. \citet[]{Hunt2019} then computed the current density of the associated FACs, and found it to be $\sim$ 5 - 10 $nA \, m^{-2}$ (Fig. \ref{fig:Bphi_FAC}), {over an order of magnitude smaller} than that of typical auroral FACs. \index{meridional current} \index{FAC}

\citet[]{Provan2019b} showed that the measured low-latitude $B_\phi$ is generally symmetric about the field-parallel point, where the spacecraft trajectory is tangent to the background field line. This symmetry is consistent with field perturbations associated with inter-hemispheric FACs {\citep[also see Fig. 3 in ][]{Khurana2018}}. \citet[]{Provan2019b} further examined the morphology of the low-latitude $B_\phi$ signature and showed that they can be categorized into four different groups: $\sim$ 35\% cases/orbits feature a single positive central peak $\sim$ 20 - 40 $nT$ (category A), $\sim$ 30\% cases feature two or three weaker positive peaks $\sim$ 10 - 20 $nT$ (category B), $\sim$ 15\% cases feature rather irregular central positive peaks in regions well inside the magnetic shell connected to the D-ring inner boundary (category C), and $\sim$ 20\% cases feature unique features including two with $\sim$ 20 - 30 $nT$ negative fields and two with $<$ 10 $nT$ fields (category U). \citet[]{Provan2019b} tried to correlate these different categories with the spacecraft altitude, local time, PPO phase, and the orbital phase of the D-68 ringlet but found no convincing correlation. \index{inter-hemispheric} \index{FAC} \index{D-ring} \index{PPO} \index{local time}

\adjustfigure{165pt}

\citet[]{Agiwal2021} explored the possibility of temporal variations in the thermospheric zonal winds as the origin of the variability observed in the low-latitude $B_\phi$. We will discuss more about this work in section \ref{sec:imp_ion}. \index{zonal wind}

\subsection{Discovery of Alfvén waves planetward of the D-ring}
\label{sec:Alfven_waves_obs}

\begin{figure*}
\begin{center}
\figurebox{5in}{}{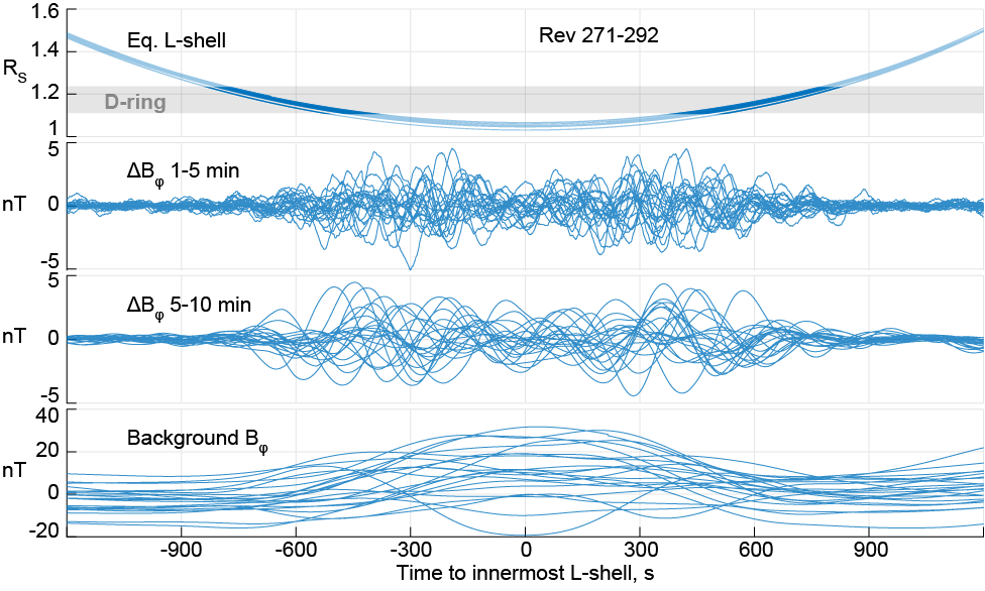}
\caption{Oscillations in the azimuthal magnetic field measured along 22 Cassini Grand Finale orbits as identified by \citet[]{Southwood2020}. The bottom panel shows the background $B_\phi$, the middle two panels show the $B_\phi$ oscillations in the 1 - 5 min band and the 5 - 10 min band, and the top panel shows the magnetic mapping of the spacecraft position to the ring plane in which the mapping to the D-ring is highlighted with the grey shade. The data is organized in the time frame in which $T=0$ represents the time when the spacecraft was on the innermost magnetic shell {(L-shell). L-shell describes the set of field lines which cross the equator at the radial distance defined by the numerical value of L.} Figure from \citet[]{Southwood2020}.}
\label{fig:Alfven_waves}
\end{center}
\end{figure*}

\citet[]{Southwood2020} examined the shorter timescale variations in the azimuthal magnetic field measured along the CGF orbits. They showed that $B_\phi$ oscillations with a typical amplitude of a few $nT$ and a typical time scale of a few minutes prevail in the innermost magnetosphere planetward of the D-ring, regardless of the morphology of the background $B_\phi$ structure. They applied cubic spline fittings with a 5-min window and a 10-min window successively to the measured $B_\phi$, which extracted two bands of $B_\phi$ oscillations (middle two panels of Fig. \ref{fig:Alfven_waves}). This procedure also isolated the large-scale background $B_\phi$, as shown in the bottom panel of Fig. \ref{fig:Alfven_waves}. It can be seen from Fig. \ref{fig:Alfven_waves} that both the large-scale background $B_\phi$ and the oscillatory $B_\phi$ are generally confined within the magnetic shell connecting to the outer edge of Saturn's D-ring. The background $B_\phi$ features a typical amplitude of 20 $nT$, while both bands of the oscillatory $B_\phi$ have amplitudes on the order of 5 $nT$ or less. \index{D-ring} \index{azimuthal magnetic field} \index{D-ring}

\begin{figure*}
\begin{center}
\vspace{129pt}
\figurebox{5in}{}{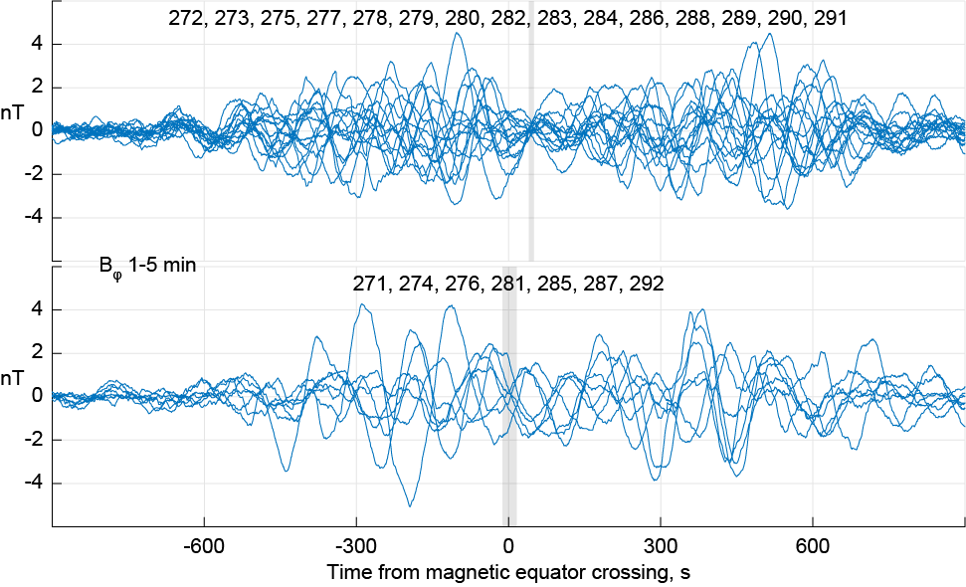}
\caption{Azimuthal magnetic field oscillations in the 1-5 min band as a function of time from the crossing of the proxy magnetic equator (where the background radial field, $B_r=0$).  Orbit Rev numbers are indicated on each panel. The top panel displays 15 CGF orbits during which a magnetic node occurred between $T=$ 40 and 50 sec (vertical shaded bar) from $B_r=0$. The lower panel displays 7 CGF orbits that show a clustering of nodes between T= $-12$ and $+16$ sec (vertical shaded bar). Figure from \citet[]{Southwood2020}.}
\label{fig:Bphi_nodes_highfreq}
\end{center}
\end{figure*}

When examining the peaks (anti-nodes) and zeros (nodes) of the $B_\phi$ oscillations with respect to the local background planetary field ($B_r$ \& $B_\theta$), \citet[]{Southwood2020} noticed that the nodes of $B_\phi$ oscillations from both bands lie close to the proxy magnetic equator (defined as where the radial component of the field vanishes $B_r=0$). For the 1-5 min oscillations, 15 orbits show a strong clustering of a $B_\phi$ node within 5 secs of $T=45$ sec from $B_r=0$; the {remaining} 7 orbits show a clustering of nodes between $T=-12$ and $+16$ sec from $B_r=0$ (Fig. \ref{fig:Bphi_nodes_highfreq}). This feature is consistent with the odd-mode (wave number $n=1, 3, 5,$ etc.) of a standing Alfvén wave (assuming similar ionospheric electrical conductivity in the northern hemisphere and southern hemisphere). For the 5-10 min oscillations, the picture is somewhat more complex: 9 orbits display a magnetic node between $T=-6$ and 35 sec on either side of the proxy magnetic equator ($B_r=0$), another 9 orbits display a magnetic node between $T=-30$ and $-40$ sec from the proxy magnetic equator. There are four additional orbits with mixed behaviors, including one with a $B_\phi$ anti-node (maximum) near the proxy magnetic equator and two with relatively flat $B_\phi$ oscillations near $B_r=0$. \index{magnetic equator} \index{node} \index{anti-node} \index{Alfvén wave}

\citet[]{Southwood2020} propose that these few minutes $B_\phi$ oscillations planetward of Saturn's D-ring are standing (transverse) Alfvén waves. The fact that 1) both oscillation bands feature a magnetic node near the proxy magnetic equator ($B_r=0$) and 2) the (temporal) spectral content of these oscillations remains relatively simple despite the up to a factor of two change in local field line length led \citet[]{Southwood2020} to suggest that these are local field line resonances being pumped by global magnetospheric cavity modes, a coupling originally suggested by \citet[]{KivelsonSouthwood1986} for the 1 - 10 min ultra‐low frequency (ULF) magnetic pulsations in Earth's magnetosphere. \index{D-ring} \index{Alfvén wave} \index{field line resonance} \index{cavity mode}

\subsection{The internal magnetic field: extreme axisymmetry and high-degree magnetic moments} \label{sec:Bint_obs}

The small amplitude of the azimuthal component ($B_\phi$ $\sim$ 0.1\% $|B|$ near {CGF} periapses) and the identifiable magnetospheric-ionospheric features in this component serve as direct evidence for the extreme axisymmetry of Saturn's internal magnetic field. Quantitative analyses confirm this first impression. Two different analyses have been performed with the CGF magnetic field measurements to quantify the amount of non-axisymmetry in Saturn's internal magnetic field \citep[]{Cao2019}: longitudinal variation in Saturn's magnetic equator {positions} and Gauss coefficients inversion with all three field components. The two analyses yielded similar upper limits on large-scale non-axisymmetry in Saturn's internal magnetic field: the dipole tilt at Saturn must be smaller than 0.007$^\circ$ (25.2 arcsecs), {while the non-axisymmetric contribution to the spherical harmonic (SH) degree 2 \& 3 magnetic moments are less than 0.15\% \citep[]{Cao2019}.} \index{non-axisymmetry} \index{magnetic moment} \index{magnetic equator} \index{dipole tilt}

Although extremely axisymmetric, Saturn's internal magnetic field displays surprisingly rich (spatial) spectral content, corresponding to an axisymmetric magnetic field on many different length-scales \citep[]{DC2018, Cao2019}. Two different mathematical representations of the internal magnetic field were adopted by \citet[]{Cao2019}: the Gauss coefficients representation and the Green's function representation. {The Green's function relates the vector magnetic field at the point of observation to its radial component at a reference surface inside the planet. For technical details of the two different representations, the interested readers can refer to the Appendices A \& B of \citet[]{Cao2019} and references therein.} \index{Green's function} \index{Gauss coefficient} \index{axisymmetric magnetic field}

Utilizing the Green's function, \citet[]{Cao2019} computed the sensitivity of the magnetic field measurements along the CGF orbits to the axisymmetric field at Saturn's ``dynamo surface", taken to be the {$a = 0.75 \, R_S, c = 0.6993 \, R_S$} isobaric surface. {Here $a$ and $c$ represent the equatorial and polar radii of the isobaric surface, respectively.} The shape of the isobaric surface was determined from interior structural models constrained by Cassini gravity measurements \citep[]{iess2019, militzer2019}. The choice for the depth of ``dynamo surface" is {guided by an estimation of where the local magnetic Reynolds number reaches order unity \citep[]{DC2018, Cao2019}.} \citet[]{Cao2019} showed that magnetic field measurements along the CGF orbits are sensitive to Saturn's large-scale {(SH degree $n \le 3 $)} axisymmetric magnetic field at depth up to very high latitude (e.g., $\pm$80$^\circ$) but likely are not very sensitive to Saturn's small-scale {($n > 3$)} axisymmetric magnetic field beyond $\pm$60$^\circ$ latitude. These sensitivity characteristics result from 1) the particular trajectory of CGF orbits with closest approach near the equator and 2) the highly axisymmetric nature of Saturn's internal magnetic field. \index{Green's function} \index{dynamo surface} \index{magnetic Reynolds number} \index{gravity} \index{axisymmetric magnetic field}

After determining the magnetodisk field \citep[]{Bunce2007} contributions orbit-by-orbit with CGF magnetic field measurements slightly away from the periapsis {and removing them from the measurements}, \citet[]{Cao2019} performed inversion analysis on the CGF magnetic field data to extract features of Saturn's internal magnetic field with both the Gauss coefficients and the Green's function representations. Inversion analysis with the Gauss coefficients representation revealed that axisymmetric Gauss coefficients up to at least {SH} degree-9 are needed to give a reasonable match to the measurements (with Root-Mean-Square residual $<5$ $nT$). {\citet[]{Cao2019} noted that, on the dynamo surface, the contribution of the internal magnetic field corresponding to the degrees 4 - 9 axisymmetric Gauss coefficients is substantially larger beyond $\pm$60$^\circ$ latitude than at lower latitudes \citep[see Fig. 12 in][]{Cao2019}. The Green’s function sensitivity analysis of the data does not support this result. \citet[]{Cao2019} employed the technique of regularized inversion, which introduces a damping parameter, $\gamma$, that regularizes the behavior of the model while simultaneously fitting the measurements \citep[for technical details, see section 5.1.2 in][]{Cao2019}.} \index{magnetodisk} \index{inversion} \index{Gauss coefficient} \index{Green's function} \index{regularized inversion}

\begin{figure}
\begin{center}
\figurebox{3.25in}{}{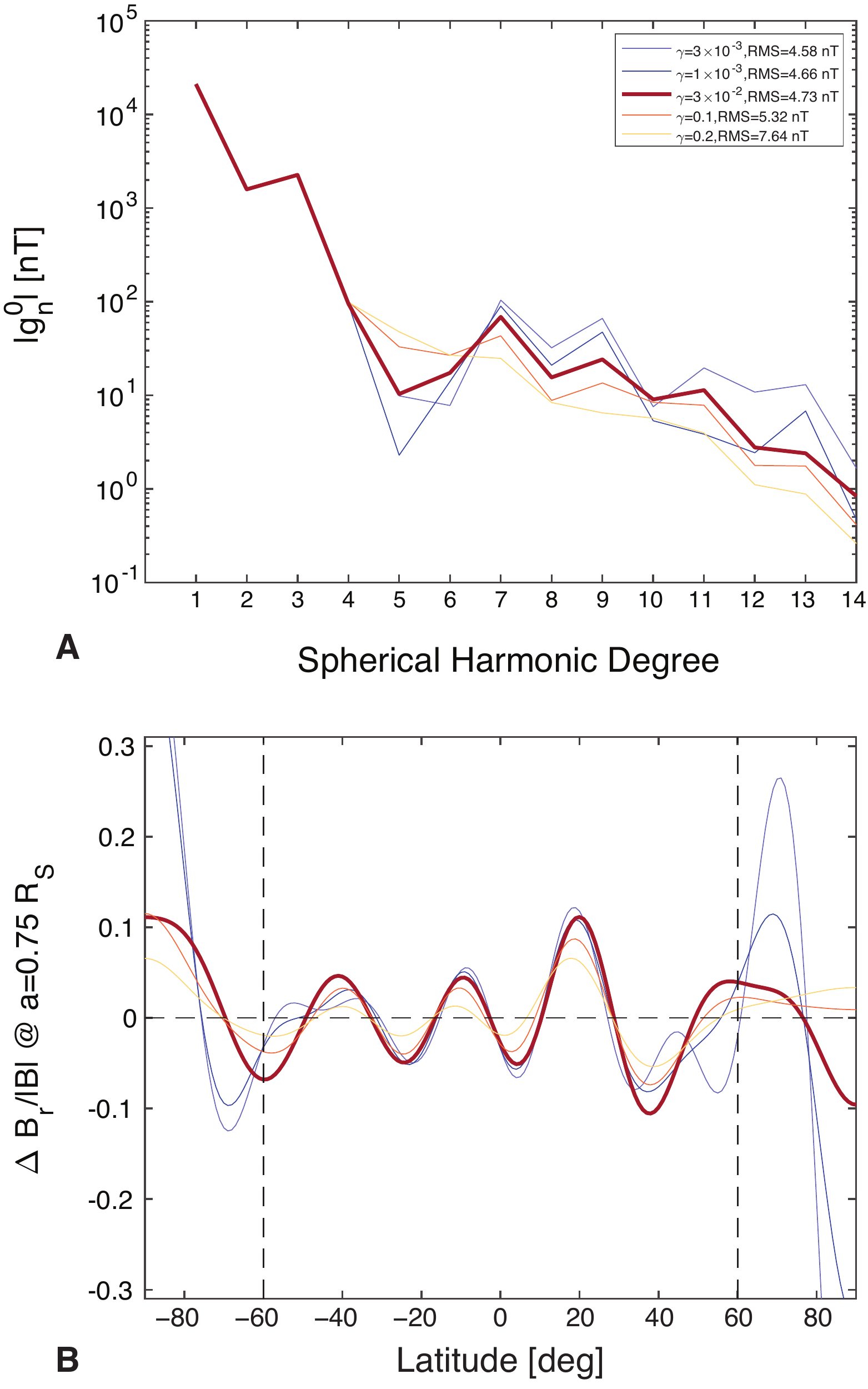}
\caption{The axisymmetric Gauss coefficients $g_n^0$ versus SH degree (panel A) and the small-scale {($g_n^0$ for $n > 3$)} internal magnetic field of Saturn versus latitude (panel B) derived from Cassini Grand Finale magnetic field measurements with regularized inversion. The damping parameter $\gamma$ and the RMS residual associated with each solution are shown in the legend in panel A. In panel B, the small-scale magnetic field, $\Delta B_r$, corresponds to $g_n^0$ with $n>3$. {It is normalized with respect to the strength of the background field, $|B|$, corresponding to $g_n^0$ with $n \le 3$.} Figure from \citet[]{Cao2019}.}
\label{fig:gn_dBr_Saturn}
\end{center}
\end{figure}

\begin{figure}
\begin{center}
\figurebox{3.5in}{}{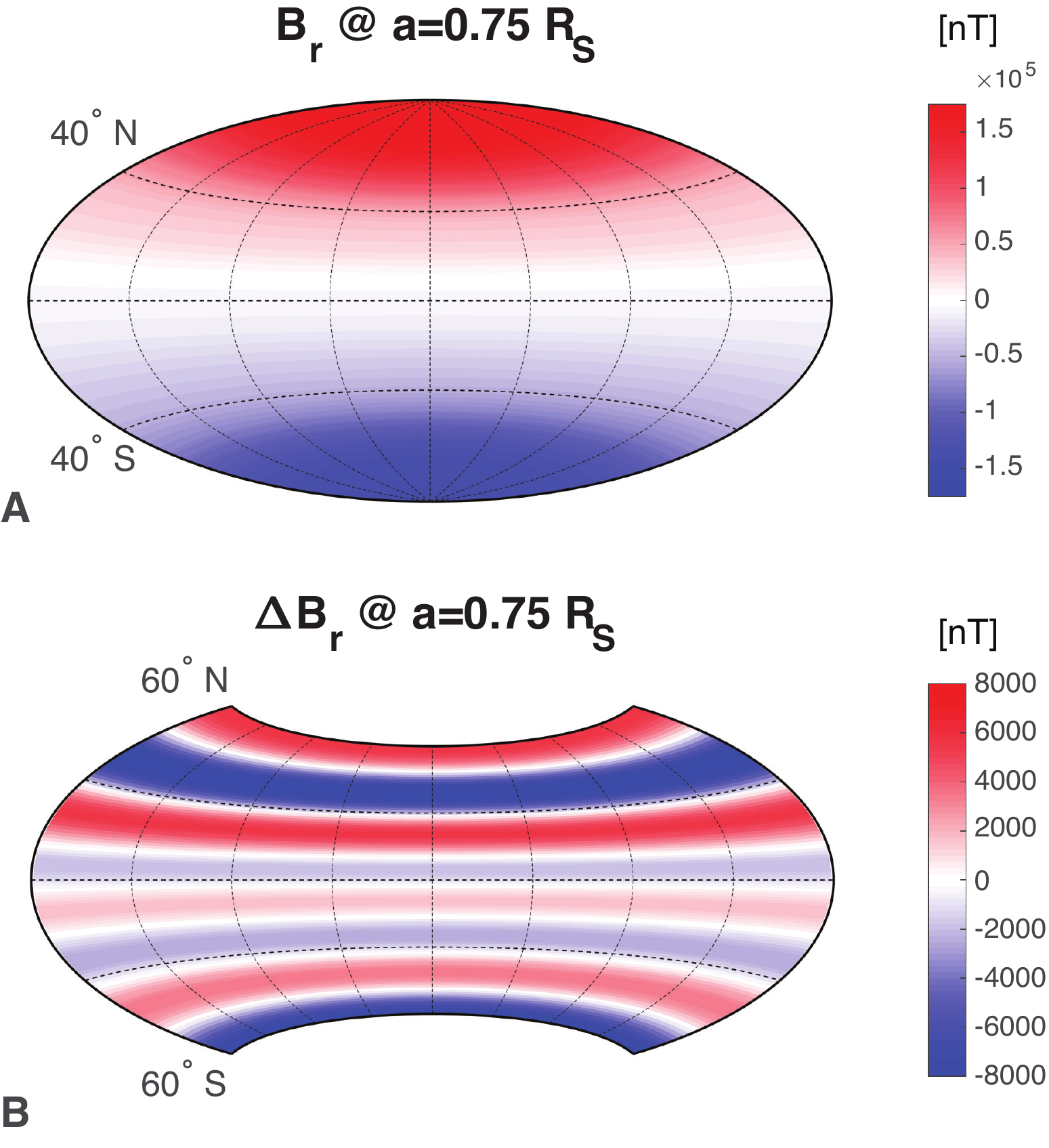}
\caption{Saturn’s large {scale ($g_n^0$ for $n \le 3$) and small scale ($g_n^0$ for $n > 3$)} radial magnetic field at the $a = 0.75$ $R_S$, $c = 0.6993$ $R_S$ isobaric surface according to the Cassini $11+$ field model. Saturn’s large scale radial magnetic field at this depth features a relatively weak equatorial region, $B_r$ remains less than 50,000 $nT$ ($<$1/3 of its peak value) between $\pm$40$^\circ$. Saturn’s small-scale magnetic field at this depth features eight alternating bands between $\pm$60$^\circ$, with typical amplitude of $\sim$ 5\%–10\% of the background field. Figure from \citet[]{Cao2019}.}
\label{fig:Br_dBr_Saturn}
\end{center}
\end{figure}

Fig. \ref{fig:gn_dBr_Saturn} shows characteristic properties of a few selected models, both in spectral space (the amplitude of $g_n^0$) and in real space ($\Delta B_r/|B|$ on the ``dynamo surface"), as a function of the damping parameter $\gamma$ {which sets the relative importance of model constraints}. {It can be seen that models tend to feature 1) large amplitude of $\Delta B_r$ at high-latitude when the damping is weak, 2) significantly reduced amplitude at high latitude when some damping is allowed, and 3) a substantially worse fit to the data when the damping is strong (see the RMS residual values in the legend in Figure \ref{fig:gn_dBr_Saturn}).} Fig. \ref{fig:gn_dBr_Saturn} further shows that the small-scale field structure within $\pm$60$^\circ$ latitude is well-resolved regardless of the field behavior at higher latitudes: there are eight alternating latitudinal bands of radial magnetic fluxes between $\pm$60$^\circ$ latitude (see also Fig. \ref{fig:Br_dBr_Saturn}B). The model that features a well-behaved field at the ``dynamo surface" and a good match to the data, corresponding to $\gamma=0.03$ (thick red traces in Fig. \ref{fig:gn_dBr_Saturn}), was selected as the preferred model and was named the Cassini $11+$ model (see Table \ref{tab:Gauss_Coeff} for the coefficients). \index{dynamo surface} \index{Cassini $11+$ model}

\citet[]{Cao2019} also performed inversion analysis on the CGF magnetic field measurements with the Green's function representation. Compared to the Cassini $11+$ field model, the Green's function analysis returned an almost identical small-scale axisymmetric magnetic field within $\pm$60$^\circ$ latitude on the ``dynamo surface", but (as expected) produced different field behaviors at higher latitudes (see their Fig. 15). \index{Green's function} \index{axisymmetric magnetic field}

With a robust understanding of which features are well-resolved by the CGF data, \citet[]{Cao2019} examined the characteristics of Saturn's internal magnetic field at the ``dynamo surface" (the $a = 0.75$ $R_S, c = 0.6993$ $R_S$ isobaric surface). As shown in Fig. \ref{fig:Br_dBr_Saturn}A, Saturn's large-scale (SH degrees 1 - 3) magnetic field {is} relatively weak {in the} equatorial region: $B_r$ remains less than 1/3 of its peak value within $\pm$40$^\circ$ latitude (panel A). As shown in Fig. \ref{fig:Br_dBr_Saturn}B, Saturn's small-scale (beyond SH degree-3) magnetic field features eight alternating zonal bands between $\pm$60$^\circ$ latitude (panel B), with typical amplitude $\sim$ 5 - 10\% of the background large-scale field (see Fig. \ref{fig:gn_dBr_Saturn}B). Like that of the large-scale field, the polarity of the eight alternating magnetic bands displays predominant anti-symmetry with respect to the equator (dipole-octupole-like), while their amplitudes are stronger at northern mid-latitudes. \citet[]{Cao2019} further noted that the number of alternating magnetic bands at the $a=0.75$ isobaric surface coincides with the number of alternating zonal wind bands if one projects the observed surface wind along the spin-axis to the same depth. Both the small-scale magnetic bands and the projected surface zonal winds feature typical latitudinal widths of $\sim$ 15$^\circ$. 
\index{dynamo surface} \index{anti-symmetry} \index{zonal wind}

{It is interesting to examine the pattern of the Gauss coefficients of Saturn’s internal field} (see Fig. \ref{fig:gn_dBr_Saturn}A and Table \ref{tab:Gauss_Coeff}). Overall, there is a dramatic drop in amplitude beyond degree-3: $g_4^0$ drops by more than an order-of-magnitude compared to $g_2^0$ and $g_3^0$, and all higher-degree moments are smaller than $g_4^0$. The degree-5 moment, $g_5^0$, seems to be strongly suppressed: it is almost another order-of-magnitude smaller compared to $g_4^0$. However, the degree-7 moment, $g_7^0$, bounces back by a factor 5 - 7 compared to $g_5^0$. $g_8^0$ to $g_{11}^0$ are comparable to $g_5^0$ and $g_6^0$. Although an overall decaying (with SH degree) trend is present, 1) the strong dip in $g_5^0$ and 2) the seemingly separate slopes for the low-degree moments and the high-degree moments are unexpected. \index{Gauss coefficient}

\citet[]{Cao2019} attempted to extract an electromagnetic induction signal from Saturn's interior, with the orbit-to-orbit varying magnetodisk $B_Z$ field as the external sounding signal. They solved for the orbit-to-orbit varying internal dipole $\Delta g_1^0$ after removing the Cassini $11+$ model, and then compared those values to the orbit-to-orbit varying magnetodisk field $\Delta B_Z$. Although the expected induction signal is within 1$\sigma$ of a formal inversion analysis of $\Delta g_1^0$ versus $\Delta B_Z$, the large scatter in the data precluded a definitive constraint on the induction depth inside Saturn. \index{electromagnetic induction} \index{interior} \index{magnetodisk}

\begin{table}
\begin{center}
\begin{tabular}{r r r}
\hline
  [$nT$] & Cassini 11 & Cassini 11+  \\
\hline
  $g_1^0$  & 21140   & 21141 \\
   $g_2^0$  & 1581  & 1583 \\
   $g_3^0$  & 2260   & 2262 \\
   $g_4^0$  & 91   & 95 \\
   $g_5^0$  & 12.6   & 10.3  \\
   $g_6^0$  & 17.2   & 17.4 \\
   $g_7^0$  & $-59.6$ &  $-68.8$ \\
   $g_8^0$  & $-10.5$ &  $-15.5$ \\
   $g_9^0$  & $-12.9$ &  $-24.2$ \\
   $g_{10}^0$  & 15   & 9.0 \\
   $g_{11}^0$  & 18   & 11.3 \\
   $g_{12}^0$  &    & $-2.8$ \\
   $g_{13}^0$  &    & $-2.4$ \\
   $g_{14}^0$  &    & $-0.8$ \\
 \hline
\end{tabular}
\caption{Gauss coefficients of the Cassini 11 model \citep[]{DC2018} and the Cassini $11+$ model \citep[]{Cao2019} for Saturn. These coefficients refer to a surface radius $R_S$=60268 $km$. The Cassini 11 model was constructed from magnetic field measurements from the first ten Cassini Grand Finale orbits while the Cassini $11+$ model was constructed using data from all 22.5 CGF orbits.}
\label{tab:Gauss_Coeff}
\end{center}
\end{table}

\section{Implications and open questions}\label{sec:implications}

Here we briefly discuss the implications and open questions for Saturn's innermost magnetosphere, ionosphere, and interior that are closely related to the magnetic field investigation. As will become clear, our understanding of the physical mechanisms behind many of the observed phenomena is still in a preliminary stage: most of our interpretations are kinematic. A fully dynamic understanding of the Saturn system {has not been obtained}. 

\subsection{Saturn's equatorial ionosphere: zonal shear and atmospheric-wave-induced temporal variability?} \label{sec:imp_ion}

\begin{figure}
\begin{center}
\figurebox{3.25in}{}{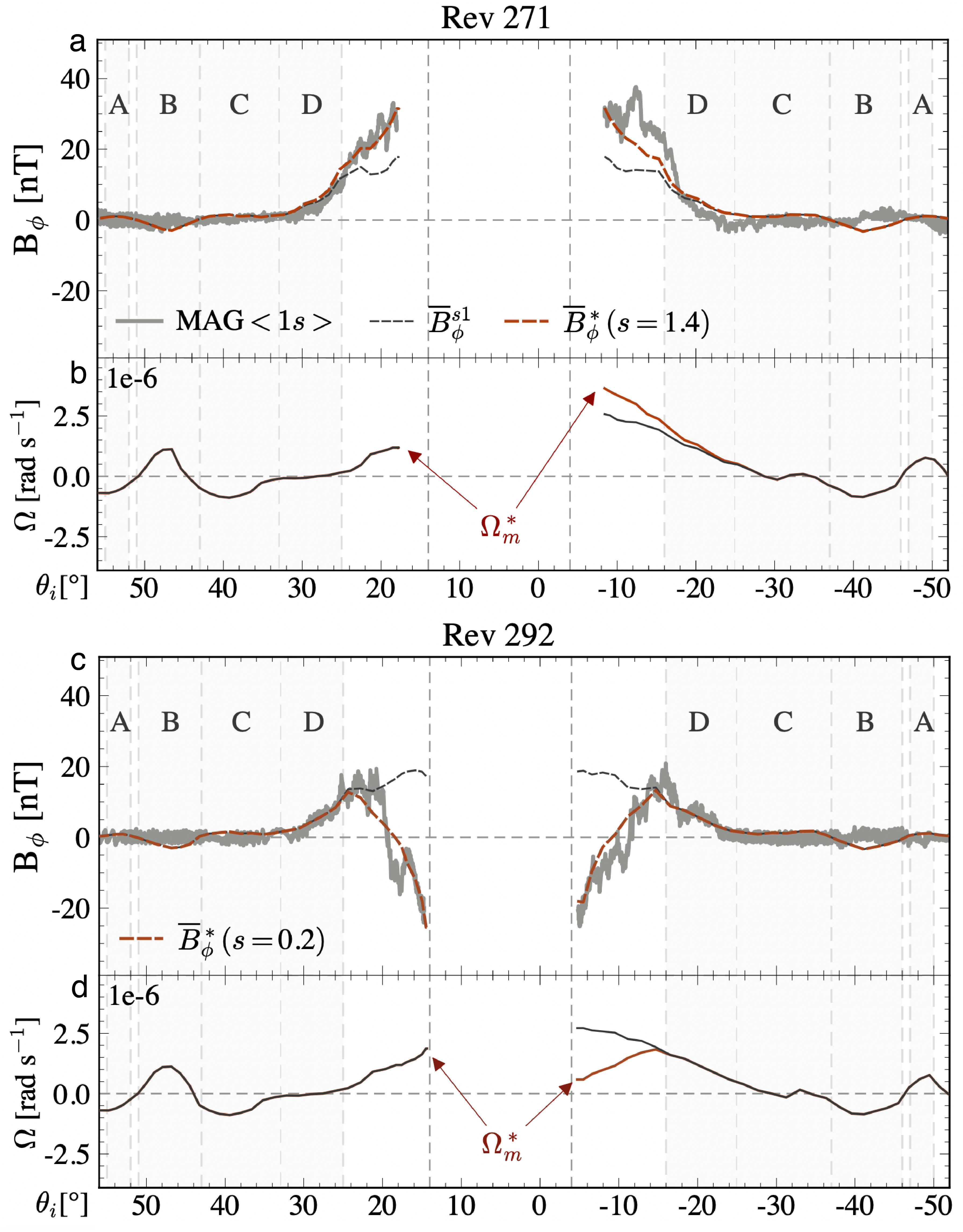}
\caption{Data-model comparisons between the observed $B_\phi$ along two CGF orbits and a kinematic ionospheric zonal shear model with different assumed zonal flow profiles. Panels a \& c display $B_\phi$, with the solid grey traces showing the de-trended observations, dashed black traces showing the model associated with the baseline zonal flow profile, and red dashed traces showing the model associated with the perturbed zonal flow profile. Panels b \& d display the angular velocity profile, with the black traces showing the baseline model, and the solid red traces showing the perturbed model. It can be seen from the data-model comparison for Rev 292 (panels c \& d) that one only needs to reduce the amplitude of the zonal flow in the southern hemisphere (instead of reversing the wind) to flip the sign of the low-latitude $B_\phi$. Figure from \citet[]{Agiwal2021}.}
\label{fig:Bphi_Variability}
\end{center}
\end{figure}

As introduced in section \ref{sec:Bphi_obs}, a considerable amount of orbit-to-orbit variability has been observed in the large-scale low-latitude $B_\phi$ peak along the Cassini Grand Finale orbits (e.g., see Fig. \ref{fig:Bphi_from_CGF}). Although most of the CGF orbits feature positive large-scale low-latitude $B_\phi$ around the periapses, a few orbits (e.g. Revs  286 \& 292) feature strong negative low-latitude $B_\phi$ (see the categorization by \citealt{Provan2019b}). As explained by \citet[]{Khurana2018}, due to the northward shift of Saturn's magnetic equator, a north-south symmetric eastward zonal flow in Saturn's ionosphere could naturally lead to a low-latitude positive $B_\phi$ (see Fig. \ref{fig:Dougherty2018_Fig0}B). \index{magnetic equator} \index{zonal flow} \index{ionosphere}

\citet[]{MW2019} and \citet[]{Brown2020} reported atmospheric waves in Saturn's thermosphere with Cassini Ion Neutral Mass Spectrometer (INMS) and Ultraviolet Imaging Spectrograph (UVIS) measurements during the Cassini Grand Finale. These atmospheric waves could assist the vertical transport of the 1-bar wind structure to higher altitude, but could also introduce temporal variability to the higher altitude winds. It is thus natural to consider time variability in the zonal shear in Saturn's ionosphere as the cause of the observed variability in the large-scale $B_\phi$ around the CGF periapses. \citet[]{Agiwal2021} investigated quantitatively the amount of variation in the ionospheric zonal wind shear needed to account for the observed variability in the low-latitude $B_\phi$. \index{thermosphere} \index{atmospheric waves} \index{temporal variability} \index{zonal wind shear} \index{ionosphere}

The technical starting point of \citet[]{Agiwal2021} is the rigorously derived formula relating the steady-state zonal shear in the ionosphere to the meridional current in the ionosphere (under a thin-shell approximation of the ionosphere) and $B_\phi$ along the field line, first presented in Appendix A of \citet[]{Provan2019b}. The steady-state meridional current, $I_m$, associated with the differential ionospheric wind drag is shown to be
\begin{equation}
    I_m=\frac{\Sigma_{PN} \Sigma_{PS} \left( \Omega_{nS}-\Omega_{nN}\right)}{\left( \Sigma_{PN}\frac{|b_{iS}|}{\rho^2_{iS}B^2_{iS}} +\Sigma_{PS}\frac{|b_{iN}|}{\rho^2_{iN}B^2_{iN}} \right)},
    \label{eqn:Im_Omega}
\end{equation}
where $\Sigma_{PN, S}$ are the height-integrated Pedersen conductivities in the northern and southern ionospheres, $\Omega_{nN,S}$ are the angular velocities of the neutral gas in the ionospheres, $B_{iN,S}$ are the planetary poloidal fields in the two ionospheres, while $b_{iN,S}$ are the poloidal field components normal to the ionosphere. Assuming axisymmetry, the associated azimuthal magnetic field at any point along the field line is
\begin{equation}
    B_\phi=\frac{\mu_0 I_m}{\rho},
    \label{eqn:Bphi_Im}
\end{equation}
where $\mu_0$ is the permeability of free-space, and $\rho$ is the cylindrical radius from the magnetic axis (also the spin-axis) of Saturn. It should be immediately clear from Eqs. \ref{eqn:Im_Omega} \& \ref{eqn:Bphi_Im} that 1) the difference between the angular velocities at the two ends of a field line, $\Omega_{nS}-\Omega_{nN}$, determines the sign of $B_\phi$, while 2) the amplitude of the height-integrated Pedersen conductivity modulates the amplitude of the $B_\phi$. These formulae also highlight the degenerate nature of the amplitude of the zonal wind shear and the ionospheric Pedersen conductivity if the magnetic field is the only observable quantity. $B_\phi$ would remain the same if we increase the zonal wind shear by a factor 2, while decreasing the height-integrated ionospheric Pedersen conductivity in both hemispheres by the same factor. \index{Pedersen conductivity} \index{angular velocity} \index{zonal wind shear} \index{meridional current} \index{ionosphere} \index{azimuthal magnetic field}

\citet[]{VRIESEMA2020} investigated the observed low-latitude $B_\phi$ at Saturn with a height resolved ionospheric-thermospheric electrodynamics model. The \citet[]{VRIESEMA2020} model also assumes steady-state and axisymmetry, and is a kinematic model. \citet[]{Agiwal2021} demonstrated the equivalence of the \citet[]{Provan2019b} formulation (Eqs. \ref{eqn:Im_Omega} \& \ref{eqn:Bphi_Im}) and the \citet[]{VRIESEMA2020} formulation in the thin ionospheric current layer approximation. 

With an ionospheric conductivity model \citep[]{MW2006, MW2019} evaluated under northern summer conditions at Saturn, \citet[]{Agiwal2021} adopted the latitudinal profile of the 1-bar atmospheric zonal winds but decreased their amplitude by a factor of two as the starting point of Saturn's low-latitude thermospheric wind. By systematically perturbing the zonal winds in both hemispheres, they showed that variability in the equatorial thermospheric wind up to 350 $m/s$ can {account for} the observed variability in the low-latitude $B_\phi$ along CGF orbits. Fig. \ref{fig:Bphi_Variability} showcases two examples from \citet[]{Agiwal2021}. The data-model comparison for Rev 292 shows that one only needs to reduce the amplitude of the zonal flow in the southern hemisphere (solid red traces in panel d), instead of reversing the flow direction, to flip the sign of $B_\phi$ around the periapses of the CGF orbits. {Future investigations that model the lower atmosphere-thermosphere-ionosphere-inner magnetosphere interactions with the relevant dynamical processes such as gravity waves are needed to evaluate whether such variability in the thermospheric wind is plausible at Saturn.} \index{northern summer} \index{zonal wind} \index{zonal flow} \index{thermosphere} \index{ionosphere}

{Furthermore, will there be a strong seasonal dependence of the thermospheric wind shear and the low-latitude field-aligned current system at Saturn? Will the low-latitude $B_\phi$ flip to mainly negative when the season switches to northern winter at Saturn?} \index{seasonal dependence} \index{field-aligned current}

\subsection{Saturn's interior: a tale of two dynamos?}
\label{sec:two_dynamos}

The fact that Saturn's internal magnetic field is extremely axisymmetric and yet full of structures in the latitudinal direction (e.g., see Fig. \ref{fig:Br_dBr_Saturn}) is intriguing. \citet[]{DC2018} and \citet[]{Cao2019} proposed that two spatially separate dynamos, one in the deep metallic hydrogen layer and the other in the semi-conducting layer, are responsible for {this} intriguing magnetic field behavior (e.g., see Fig. \ref{fig:Dougherty2018_Fig0}A). The deep dynamo is responsible for generating the large-scale dipolar background field, while the shallow secondary dynamo \citep[]{cao2017b} is responsible for generating the small-scale latitudinally banded magnetic perturbations. \index{dynamo} \index{metallic hydrogen} \index{semi-conducting} \index{deep dynamo} \index{secondary dynamo} \index{interior}

{\citet[]{gastine2014} observed localized secondary dynamo action at low-latitudes in their 3D Jovian dynamo simulations which they attributed to interaction with deep zonal winds.}
\citet[]{cao2017b} proposed the possibility of a {global} secondary dynamo inside Jupiter and Saturn, and illustrated this process with mean-field electrodynamics \citep[]{moffatt_dormy_2019}. Three key ingredients of their envisioned secondary dynamo inside giant planets are: 1) the background magnetic field, $\mathbf{B_0}$, generated by the deep dynamo; 2) differential rotation (zonal flows) in the semi-conducting layer which generates the toroidal magnetic field, $\mathbf{B_T}$, via the $\omega-$effect (this process bears similarity to the proposed process of zonal wind shear in the ionosphere generating the low-latitude $B_\phi$); and 3) small-scale helical convection in the semi-conducting layer providing the critical $\alpha-$effect that generates the externally observable poloidal magnetic field, $\Delta \mathbf{B_P}$, from $\mathbf{B_T}$. Both $\mathbf{B_T}$ and $\Delta \mathbf{B_P}$ are expected to be spatially correlated with the zonal flows in the semi-conducting layer. {Furthermore, \citet[]{cao2017b} pointed out that the $\omega-$effect and the $\alpha-$effect in the secondary dynamo may operate at different depths: $\mathbf{B_T}$ resulting from the $\omega-$effect can be generated at a relatively shallow depth and then diffuse several scale-heights downward due to the rapidly increasing electrical conductivity as a function of depth in the semi-conducting layer.} \index{secondary dynamo} \index{zonal wind} \index{zonal flow} \index{mean-field electrodynamics} \index{secondary dynamo} \index{deep dynamo} \index{zonal wind shear} \index{omega-effect} \index{alpha-effect} \index{semi-conducting} \index{differential rotation}

\citet[]{GalantiKaspi2020} analyzed Saturn's gravity moments and magnetic moments jointly, aiming to construct a deep differential rotation profile for Saturn. Their analysis of Saturn's dynamic gravity moments was based on the Cassini radio tracking observations \citep[]{iess2019} and the diagnostic thermal wind {(TW)} relation \citep[]{kaspi2013, CS2017JGR}. Their analysis of Saturn's magnetic moments was based on the Cassini 11 model \citep[]{DC2018} and the (kinematic) mean-field electrodynamics {(MFED)} model of \citet[]{cao2017b}. They showed that the same deep zonal wind profile, which is only slightly modified from the observed surface zonal winds around $\pm$30$^\circ$ latitude, {and} penetrating to about 7500 $km$ below the 1-bar level can account for both the dynamic gravity field and the small-scale magnetic bands at Saturn. This is an encouraging result, and calls for future dynamic investigations, {in which the physical processes that control the depth of rapid zonal flows inside Saturn (and Jupiter) need to be self-consistently modeled.} \index{gravity} \index{thermal wind} \index{zonal wind} \index{magnetic moment}

\subsection{{Saturn’s extremely axisymmetric magnetic field: electromagnetic filtering or a double-diffusive dynamo in the diffusive core?}} \label{sec:numerical_dynamos}

The axisymmetry of Saturn's internal magnetic field has long been a puzzle. Here we briefly explain the challenge and then describe the proposed kinematic solution and some dynamical tests with 3D numerical dynamo simulations. {We conclude this subsection with a discussion of the new challenges raised by the inference of a large diffusive core inside Saturn \citep[]{MankovichFuller2021}.} \index{dynamo} \index{diffusive core}

From a theoretical point of view, Cowling's theorem \citep[]{Cowling1933} precludes the possibility of maintaining a purely axisymmetric magnetic field in a steady-state via MHD dynamos. However, Cowling's theorem itself does not place any lower limit on the amount of non-axisymmetry necessary to maintain dynamo action. When examining the observational evidence in the solar system, Earth and Jupiter feature a modest amount of non-axisymmetry with dipole tilts $\sim$ 10$^\circ$, Uranus and Neptune feature a significant amount of non-axisymmetry with dipole tilts $\sim$ 50$^\circ$ \citep[e.g., see Table 7.2 in][]{KivelsonBagenal2014}. {For Mercury and Ganymede, the total amount of non-axisymmetry in their large-scale internal magnetic field are less clear at this stage while their dipole tilts have been estimated to be on the order of 1$^\circ$.} \index{Cowling's theorem} \index{dynamo} \index{dipole tilt}

The most widely invoked explanation for Saturn's very axisymmetric magnetic field is the one proposed by \citet[]{Stevenson1980, Stevenson1982}. The essence of this mechanism is electromagnetic (EM) filtering: if there exists a passive, electrically conducting layer on top of a regular dynamo, and if this filter layer is rotating at a different angular speed compared to that of the deep dynamo, then any non-axisymmetric magnetic field from the deep dynamo would appear as a time-varying field to this layer and thus be electromagnetic filtered while the axisymmetric part of the magnetic field would appear as time-stationary and pass through. For Saturn in particular, \citet[]{Stevenson1980} proposed that helium rain-out would create a stably stratified layer (with no large-scale overturning motion) on top of Saturn's deep dynamo. One {should remember} that both the qualitative description above and the quantitative model presented in \citet[]{Stevenson1982} are kinematic: no dynamical feedback between the deep dynamo and the stably stratified layer has been considered. Within this kinematic framework, \citet[]{Cao2019} showed that a 0.007$^\circ$ dipole tilt requires a stable layer thicker than 2500 $km$ if the deep dynamo features a Jupiter-like $\sim$ 10$^\circ$ dipole tilt. This estimation represents a lower limit on the thickness of the stable layer, as the dynamical feedback from the non-axisymmetric magnetic field on the flows in the stable layer is expected to reduce the filtering efficiency. \index{electromagnetic filtering} \index{helium rain} \index{stably stratified} \index{stable layer}

3D numerical dynamo models have also been constructed to reproduce the highly axisymmetric magnetic field of Saturn. \citet[]{CW2008,Stanley2008,Stanley2010,Christensen2018Geo,Gastine2020, YanStanley2021} employed numerical MHD models under the Boussinesq approximation to investigate the effects of a stable layer on top of a convective dynamo. In these models, background density and electrical conductivity are assumed to be constant while the Ohmic and viscous dissipation are assumed to be negligible. While these assumptions are approximately valid for the cores of terrestrial planets, they are quite different from the conditions inside giant planets, which feature large variations in background density, temperature, and electrical conductivity {as a function of depth} \citep[]{french2012}. \index{Boussinesq approximation} 

These simplified models illustrate many dynamical processes involved in the interaction between the convective dynamo and the stable layer. The electromagnetic filtering effect proposed by \citet{Stevenson1982} was indeed observed in many of these studies \citep[e.g.][]{CW2008, Stanley2010, Gastine2020}. However, the situation is more complicated than the kinematic picture. Convection eddies can penetrate into the stably stratified layer \citep[]{Gastine2020}, horizontal circulation can erase {certain types of} thermal heterogeneity imposed from above \citep[]{Christensen2018Geo}, certain types of zonal flow in the stable layer could destabilize the dynamo-generated magnetic field instead of axisymmetrizing it \citep[]{Stanley2008}. {In one of the latest 3D Boussinesq modeling effort, \citet[]{YanStanley2021} achieved a dipole tilt $\sim$ 0.066$^\circ$ with a thick stably stratified layer ($\sim$ 0.28 $R_S$) combined with a particular type of heterogeneous heat flux variation on top of the stable layer. This is still about one order of magnitude larger than the latest observational upper bound at Saturn \citep[]{DC2018,Cao2019}. Part of this discrepancy could result from the relatively low magnetic Reynolds number associated with differential rotation in the stable layer in the numerical simulations compared to the realistic values inside Saturn. See section 4.4.3 in \citet[]{ChristensenCao2018} for a more detailed discussion of this point.} \index{dynamo} \index{stable layer} \index{electromagnetic filtering} \index{zonal flow} \index{magnetic Reynolds number} 

More advanced models, where the molecular envelope and the deep dynamo with smooth transition in material properties were simulated simultaneously, have also been constructed under the anelastic approximation \citep{jones2014, DJ2018, GastineWicht2021}. Most of the anelastic dynamo models were constructed without any stable layer \citep[]{Duarte2013, jones2014, DJ2018}, the resulting magnetic field in the dipolar branch features a modest amount of non-axisymmetry similar to that observed at Jupiter. In a survey of the effects of relative thickness of the molecular envelope, \citet{DJ2018} observed an interesting oscillatory dynamo in which the magnetic field flips its polarity regularly and resembles Saturn's main magnetic field qualitatively during about a quarter of every cycle. The dipole tilts during such times are typically 1-2$^\circ$. \index{anelastic approximation} \index{dipole tilt}

{Three} recent studies implemented stable layers in anelastic models applicable to Jupiter and Saturn. \citet{dietrich2018} constructed a hydrodynamic model for Saturn with a sandwiched stable layer and analyzed the depth and mechanism of penetrative convection. \citet{Christensen2020} investigated how a stable layer might work with the dynamo-generated magnetic field to truncate deep zonal flows in a 2.5D (axisymmetric) set-up. {\citet{GastineWicht2021} constructed one of the first 3D MHD model for gas giants with a stably stratified layer between 0.82 $R_P$ and 0.86 $R_P$ near the molecular-metallic transition. The resultant surface magnetic field are dipole-dominant with appreciable amount of non-axisymmetry (see their Fig. 9), similar to that of Jupiter.} \index{stable layer} \index{zonal flow} \index{non-axisymmetry}

{The latest inference of a large stably stratified diffusive core inside Saturn \citep[]{MankovichFuller2021} raised new challenges for understanding the origin of Saturn's axisymmetric magnetic field. A diffusive core extending out to 0.6 $R_S$ (see Fig. \ref{fig:SaturnInterior}BC) leaves very little space for a traditional convective dynamo inside Saturn. Can an MHD dynamo operate in a stably stratified diffusive core? Small-scale double diffusive or oscillatory convection are expected to exist in the stably stratified layer(s) inside Saturn. Could these small-scale motions produce the necessary $\alpha-$effect to maintain an MHD dynamo and generate a highly axisymmetric magnetic field?} \index{stably stratified} \index{diffusive core} \index{dynamo} \index{double diffusive convection} \index{alpha-effect} \index{axisymmetric magnetic field}

\section{Outlook for future exploration of Saturn and other giant planets} \label{sec:outlook}

It is appropriate to end this data-driven chapter with an outlook for future exploration of Saturn and other giant planets, focusing on aspects related to magnetic field investigations. In the Saturn system, other than the intriguing moons Enceladus and Titan, four areas near Saturn that have not been visited in-situ are 1) the low altitude region near the poles of Saturn, 2) regions inside the orbit of F-ring at local time sectors away from noon, 3) the equatorial region immediately above and below the rings, and 4) the atmosphere of Saturn. Magnetic field measurements in these four areas will help 1) resolve Saturn's small-scale magnetic field, including any local non-axisymmetry, near the poles, 2) map out EM coupling between Saturn's northern and southern ionospheres as well as between Saturn and its rings at different local time sectors, 3) clarify the existence of a ``ring ionosphere", and 4) distinguish the ionospheric magnetic field and the internal magnetic field from the deep interior. \index{magnetic field investigation} \index{local time} \index{non-axisymmetry} \index{ionosphere}

Comparing different planets is an important step to differentiate common processes from special realizations. After the Juno mapping campaign at Jupiter, we shall have a much better knowledge of the similarities and differences between Jupiter and Saturn. Orbital missions to Uranus and Neptune are needed to complete our mapping of the Solar System giant planets. The single Voyager 2 flyby at Uranus and Neptune informed us that Uranus and Neptune possess fundamentally different magnetic fields compared to those at Jupiter and Saturn {\citep[see a recent review by][]{SS2020}}. However, {meaningful theoretical interpretations of the Uranus and Neptune systems will require significantly more constraints from in situ observations.}
\index{Voyager} \index{Uranus} \index{Neptune}

The ever expanding categories of exoplanets offers many giant planets for investigation, some of which function {in} extreme environments (e.g., the hot Jupiters). Measuring their surface composition, wind pattern, and magnetic field can provide unique observational constraints {in} vastly different settings compared to giant planets in the solar system. As in many other circumstances, studying the extremes could help us understand the ``norm". Last but not least, measurements and theoretical studies at the giant planets also help us better understand planet Earth, as many processes are common {despite the differences in the parameter regimes in which they function}. 

\section*{Acknowledgement}
H.C. is funded by the NASA Cassini Data Analysis Program (Grant Number 80NSSC21K1128).  H.C.'s visit to Imperial College London was funded by the Royal Society, UK grant RP 180014. Work at Imperial College London was funded by Science and Technology Facilities Council (STFC), UK consolidated grant ST/N000692/1. M.K.D. is funded by Royal Society, UK Research Professorship RP140004. Work at the University of Leicester was funded by STFC, UK consolidated grant ST/N000749/1. E.J.B. was supported by a Royal Society Wolfson Research Merit Award.

\bibliography{Cao2021SatMag}\label{refs}
\bibliographystyle{cambridgeauthordate-Leigh}


 \printindex

\end{document}